# Self-consistent potential correction for charged periodic systems

Mauricio Chagas da Silva,[1,2,3,†] Michael Lorke,[1] Bálint Aradi,[1] Meisam Farzalipour Tabriz,[1,4] Thomas Frauenheim,[1,5] Angel Rubio,[2,6] Dario Rocca,[3,†] and Peter Deák[1,†]

[1] Bremen Center for Computational Materials Science, University of Bremen, P.O. Box 330440, D-28334 Bremen, Germany.

[2] Max Planck Institute for the Structure and Dynamics of Matter; Luruper Chaussee 149, Geb. 99 (CFEL), 22761 Hamburg, Germany.

[3] Université de Lorraine, CNRS, LPCT, F-54000 Nancy, France.

[4] Max Planck Computing and Data Facility, Gießenbachstr. 2, D-85748 Garching, Germany

[5] Computational Science Research Center, No.10 East Xibeiwang Road, Beijing 100193 and Computational Science and Applied Research Institute Shenzhen, China.

[6] Nano-Bio Spectroscopy Group, Departamento de Fisica de Materiales, Universidad del País Vasco UPV/EHU- 20018 San Sebastián, Spain

**Abstract**

Supercell models are often used to calculate the electronic structure of local deviations from the ideal periodicity in the bulk or on the surface of a crystal or in wires. When the defect or adsorbent is charged, a jellium counter charge is applied to maintain overall neutrality, but the interaction of the artificially repeated charges has to be corrected, both in the total energy and in the one-electron eigenvalues and eigenstates. This becomes paramount in slab or wire calculations, where the jellium counter charge may induce spurious states in the vacuum. We present here a self-consistent potential correction scheme and provide successful tests of it for bulk and slab calculations.

**Keywords:** charge correction in periodic systems; electronic properties of defects, photocatalysis, negative electron affinity surfaces, image states, ghost states.

† **Corresponding authors:**
 *mauricio.chagas-da-silva@univ-lorraine.fr*
*dario.rocca@univ-lorraine.fr*
*peter.deak@bccms.uni-bremen.de*

Electronic structure calculations in periodic systems are often carried out in a supercell model, i.e., for a multiple of the primitive unit, which contains and also repeats any local deviation from periodicity, as, e.g., point defects or adsorbates.[1,2] In many cases, like modeling charge assisted surface reactions to understand photocatalytic processes, or calculating charge transition levels to establish the electronic activity of a defect in a semiconductor, the supercell may contain a localized extra charge. To avoid the divergence of the Coulomb-energy due to repetition of the supercell, a background charge of opposite sign is always introduced to keep the supercell neutral. However, because of the spurious interaction between repeated charges, corrections are needed.[3,4,5] This was partially solved by Coulomb cut-off techniques[6] or *a posteriori* total energy correction schemes, both for bulk[7,8,9,10,11,12] and lower dimensional systems.[13,14,15,16,17,18,19] It has been pointed out, however, that the artificial repetition of the charge affects also the localized one-electron levels.[20] Actually, correction of the electrostatic potential is required and a self-consistent potential correction scheme has been proposed for cubic bulk systems.[21] Self-consistent potential correction for low-dimensional periodic systems is difficult because of the variation in the dielectric profile at the interface, and the methods suggested so far are restricted to special cases.[22,23,24]

In low-dimensional charged periodic systems another problem arises. Except for the virtual crystal approach, where the charge of the nuclei is artificially modified,[25,26] and a recently published work,[24] which avoids the use of a jellium by calculating an excited state, the counter charge is distributed evenly in the model. Therefore, a substantial part of the countercharge is in the vacuum region. As we will demonstrate, in case of negative electron affinity surfaces, or when the thickness of the vacuum region, separating the repeated solid parts, is sufficiently wide to decouple the repeated interfaces, this leads to artificial bound states in the middle of the vacuum region. In plane-wave based calculations these states may even be erroneously occupied, but this artificial electrostatic potential influences also the results of localized basis calculations. To our knowledge, this phenomenon has been reported so far in one paper only,[27] but was not treated in a self-consistent manner.

In this letter we introduce a general, self-consistent potential correction (SCPC) method for charged periodic systems, equally applicable to wires (1D), slabs (2D) and bulk (3D) systems. The method has been implemented into the VASP electronic structure package,[28] and is available for certified users as a patch. We show that the total energy converges similarly as in the previous *a posteriori* energy correction schemes, the one electron energies are corrected automatically, and the artificial ghost states in low dimensional systems disappear.

The SCPC method applies a corrective potential ($V_{cor}$), included in the Kohn-Sham (KS) equations. During the iterative solution of the latter, $V_{cor}$ is updated according to the following four steps: *i)* the distribution of the extra charge in the supercell ($\delta\rho$) is determined, *ii)* the corresponding periodic electrostatic potential ($V_{per}$) is calculated, *iii)* the potential for the same but isolated charge distribution ($V_{iso}$) is determined by using open (Dirichlet) boundary conditions, and finally, *iv)* $V_{per}$ and $V_{iso}$ are used to determine the corrective potential $V_{corr}$, which is added to the total electronic potential. It should be noted that SCPC always aligns the final potential, considering the difference between the electrostatic potentials of the charged and the reference system far away from the defect position (in the spirit of the method of Lany and Zunger)[29].

Earlier energy correction methods[7,8,9,15,16,17,18,30] are typically based on approximate model representations of the defect charge density $\delta\rho$, such as point charges or Gaussian functions. The parameters of the model are tuned to reproduce the potential from the electronic structure calculation. Repeating the fitting procedure during the SCF is impractical so, in SCPC, $\delta\rho$ is built directly from the difference between the electronic density of the charged defect system ($\rho^{chg}$) and of the reference system ($\rho^{ref}$) on a real space grid:

$$\delta\rho(r) = \rho^{chg}(r) - \rho^{ref}(r) \quad (1)$$

which is renormalized in each iteration to the nominal value of the defect charge.

The reference system can either be the pristine system (unperturbed supercell), as in the FNV (Freysoldt – Neugebauer - Van de Walle) method,[9] or the neutral defect (at the same geometry as the charged one), as in SLABCC (slab charge correction).[18] The latter is, in principle, the correct procedure,[15] but the former seems to be more practical. Using the pristine system as reference for a self-consistent correction is especially appealing because then $\rho^{ref}$ does not have to be updated. In the examples below, unrelaxed defects have been used and, as will be shown, using either the pristine system or the neutral defect does not make much of a difference in these cases. At the relaxed geometry of the defect, however, the pristine system cannot be applied, since the changing position of atoms over the whole supercell would create artefacts in $\delta\rho$. (Note that in case of the FNV method the use of a single Gaussian model for $\delta\rho$ ignores those,[31] but SCPC is using a more realistic $\delta\rho$). Of course, using the neutral defect as reference means the need for updating the reference density. Our preliminary experience shows that the update is not necessary at each geometry optimization step. The required frequency of updates depends on the nature of the defect. The success of *a posteriori* correction methods indicates that, in case of deep-level bulk defects, SCPC need only to be applied for the final geometry. On the other hand, in slabs with ghost states, SCPC must be used during the geometry optimization process with frequent update of the reference.

When $\delta\rho$ is updated during the self-consistent procedure, the corresponding periodic electrostatic potential ($V_{per}$) is obtained by solving the Poisson equation

$$\nabla[\varepsilon(\boldsymbol{r})\nabla V_{per}(\boldsymbol{r})] = -4\pi(\delta\rho(\boldsymbol{r}) - \langle\delta\rho\rangle). \tag{2}$$

The difference between the periodic electrostatic potential of the charged defect system ($V_{chg}$) and of the reference system ($V_{ref}$) on the supercell edges is used to determine the Dirichlet boundary conditions to solve Eq.(2), using the DL_MG[32,33] open source library. The dielectric profile of the material ($\varepsilon$) is set to a homogeneous constant for charged defects in bulk systems or to a (smoothened) boxcar shape function (see the Supplemental Material: *Dielectric function in slabs*, Fig.S3) for defects in slab models, as in Ref.[15].

To determine $V_{iso}$ for the isolated defect charge, a self-consistent process was implemented to incorporate the macroscopic dielectric profile of the material in the calculation with open-boundary conditions, following Fisicaro *et al.*[34]

$$\nabla^2 V_{iso}(\boldsymbol{r}) = -4\pi\left[\frac{\delta\rho(r)}{\varepsilon(r)} + \rho^{iter}(\boldsymbol{r})\right] \tag{3}$$

with

$$\rho^{iter}(\boldsymbol{r}) = \frac{1}{4\pi}\nabla ln\varepsilon(\boldsymbol{r}) \cdot \nabla V_{iso}(\boldsymbol{r}) \tag{4}$$

The numerical solution of Eq.(3) is obtained using the open-source library PSPFFT.[35] For the bulk case, the solution is simple, since $\varepsilon(r)$ is a constant. However, we used the iterative algorithm to obtain a reasonable approximation of $V_{iso}$ also for cases with an inhomogeneous $\varepsilon$.

Finally, $V_{cor}$ is obtained by solving the Poisson equation for the compensating jellium background[4,36]

$$\nabla[\varepsilon(\boldsymbol{r})\nabla V_{cor}(\boldsymbol{r})] = -4\pi\langle\delta\rho\rangle \tag{5}$$

using the difference between $V_{iso}$ and $V_{per}$ at the edges of the supercell to determine the Dirichlet boundary conditions. While computing the difference $V_{cor} = V_{iso} - V_{per}$ directly leads to the same result, the solution of Eq.(5) presents some practical advantage. For getting the difference, separate evaluation of the potentials $V_{iso}$ and $V_{per}$ would be needed on a fine grid, and this task is particularly expensive for Eqs.(3-4). The potential $V_{cor}$ is a smooth function and the numerical solution of Eq.(5) can be obtained on a rather coarse grid; in this case the computation of $V_{iso}$ is required only at the boundaries.[36]

It is important to notice that, in practice, the corrective potential $V_{cor}$ compensates the spurious effect of the periodicity with a model potential $V_{per}$ and replaces its contribution with the model potential $V_{iso}$ of a single isolated defect. When employed within periodic calculations, the open boundary condition potential $V_{iso}$ approaches the edges of the supercell with non-zero derivative (see Fig.S1). This non-smooth behavior might lead to an undesirable electronic density redistribution within a material during a self-consistent procedure. In practice, this effect is typically limited and does not significantly affect the final results (e.g. formation energies). In order to prevent possible instabilities during iteration, a damping procedure on the charge at the boundaries can be applied in our practical implementation (see Supporting Material: *Damping region at the boundary*). Since this inconsistency between periodic and open boundary conditions decreases by increasing the cell size, potentially pathological situations can be systematically detected from convergence tests (as described for the case of $V_C$(2-) in diamond in Table.S1 and Fig.S2.) Eventually, damping is not always needed, but it helps the convergence in many cases.

Unlike commonly used *a posteriori* correction schemes for the *total energy*, the SCPC method corrects in a seamless way the potential, and also the charge density and the eigenvalues. As will be shown, the effect of the corrective potential is particularly significant in slab models where unphysical behavior occurs due to the compensating charge.

For testing the SCPC method, the convergence of the formation energy with system size was monitored and compared to the results obtained from *a posteriori* energy correction methods for a doubly negative vacancy $V_C$(2-) and the negative (nitrogen + vacancy) center $N_CV_C$(-) in bulk diamond, as well as for a singly positive chlorine vacancy $V_{Cl}$(+) in bulk NaCl and both on the surface and at the middle of a NaCl (001) slab. The effect of SCPC on ghost states, which arise in the vacuum part of a slab model due to the counter charge, was checked in a hydrogen-saturated (001) diamond slab containing the $V_CN_C$(-) center in the middle,[37] and in an anatase-TiO$_2$ (101) slab with an adsorbed O$_2$(-) molecule on the surface. All calculations were carried out with the Vienna Ab-initio Simulation Package (VASP),[28] using the projector augmented wave (PAW) method,[38] and the Perdew, Burke and Ernzerhof (PBE) exchange-correlation functional.[39] Details, including the supercell sizes, are given in the Supplemental Material: *Details of the calculations*.

As an example in the bulk, Table 1 (and Fig.S4) shows the size convergence of the formation energy in the case of $V_C$(2-) in diamond. The convergent SCPC value is within ~0.1 eV of the result obtained by the *a posteriori* method with the same reference. Note that the FNV method[9] uses the pristine system, while SLABCC,[18] which implements the scheme of Komsa et al.,[15,16] uses the neutral defect (at the same geometry as the charged one) as reference. The convergent SCPC formation energies are consequently somewhat lower than the results with *a posteriori* correction. The difference reflects the effect of self-consistency. Using the result of the first iteration in SCPC (starting from the uncorrected wave function), it is possible to recover the *a posteriori energy* correction. As shown in Table 1, these non-self-consistent SCPC results agree very well with the *a posteriori* corrections of the FNV and SLABCC methods. Fig.S5 in the Supplemental Material compares the size-dependence of the gap level of $V_C$(2-), obtained by SCPC, to the value given by the *a posteriori* correction of Chen and Pasquarello,[21] $\varepsilon_{corr} = -2E_{corr}^{tot}/2q$ (using the FNV total energy correction). The SCPC and FNV curves are very similar, proving that the defect levels are also corrected automatically in SCPC. At the largest supercell size the SCPC gap level is 0.08 eV lower, which makes for most of the 0.12 eV difference in the total energy (see Table 1).

As additional tests in bulk material, we have investigated the size-convergence of the formation energy for the $N_C V_C$(-) center in diamond (Fig.S6 and Table.S2) and for the $V_{Cl}$(+) defect in the bulk of the ionic compound NaCl (Fig.S7 and Table.S3). In both cases, the non-self-consistent SCPC results reproduce the *a posteriori* corrections. The effect of self-consistency is smallest for $N_C V_C$(-), which is a very compact, deep-level (well-localized) defect.

**Table 1.** The formation energy of $V_C$(2-) for different diamond supercell sizes, without correction (default), with *a posteriori* energy corrections (FNV[9] and SLABCC[18]), and with SCPC. SCPC-1 and -2 correspond to using the neutral defect and the pristine system, respectively, as reference. Formation energies are given in eV and the length of the cubic supercell in Å.

| Size (Å) | DEFAULT | FNV | SCPC-2 | | SLABCC | SCPC-1 | |
|---|---|---|---|---|---|---|---|
| | | | Non-SC | SC | | Non-SC | SC |
| 7.14 | 10.60 | 12.98 | 13.04 | 12.89 | - | 12.24 | 12.13 |
| 10.71 | 11.79 | 13.21 | 13.23 | 13.13 | 13.11 | 12.94 | 12.87 |
| 14.29 | 12.16 | 13.19 | 13.19 | 13.13 | 13.19 | 13.05 | 13.01 |
| 17.86 | 12.38 | 13.18 | 13.18 | 13.13 | 13.17 | 13.09 | 13.05 |

Table 2 (and Fig.S8) shows the formation energy of $V_{Cl}$(+) at the surface of a NaCl (001) slab, as a function of the vacuum thickness. The first step in the iteration with SCPC reproduces the results of the *a posteriori* corrections here as well, while the self-consistent procedure results in somewhat lower formation energies, just as in the bulk case. For the sake of completeness, we have also checked the convergence with the lateral size (Fig.S9, Table.S4) and with the thickness (Fig.S10, Table.S5) of the slab. We also checked the convergence with vacuum thickness for $V_{Cl}(+)$ at the middle layer of the slab (Fig.S11 and Table.S6). The convergence was satisfactory in all cases, and the non-self-consistent correction reproduced the *a posteriori* results well.

**Table 2:** Formation energy of $V_{Cl}(+)$ on the surface of a NaCl (001) slab as a function of vacuum thicknesses, without correction (default), with *a posteriori* energy corrections (FN[17] and SLABCC[18]), and with SCPC. Formation energies are given in eV.

| Vacuum thickness (Å) | DEFAULT | FNV | SCPC-2 | | SLABCC | SCPC-1 | |
|---|---|---|---|---|---|---|---|
| | | | Non-SC | SC | | Non-SC | SC |
| 23 | 1.60 | 1.95 | 1.92 | 1.83 | 1.89 | 1.87 | 1.77 |
| 34 | 1.81 | 1.95 | 1.92 | 1.81 | 1.89 | 1.88 | 1.76 |
| 45 | 2.04 | 1.95 | 1.92 | 1.80 | 1.89 | 1.91 | 1.76 |
| 68 | 2.55 | 1.95 | 1.92 | 1.79 | 1.89 | 1.94 | 1.77 |

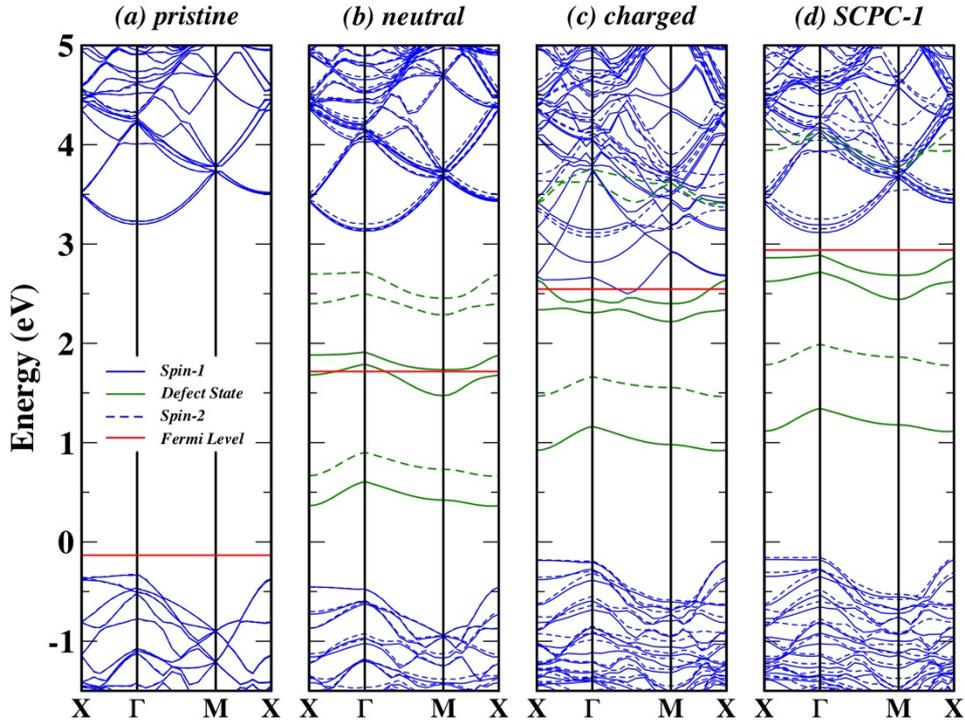

**Fig.1.** The band structure of a hydrogen saturated diamond (001) slab, without (a) and with (b-d) an $N_CV_C$ center in the middle of the slab. The neutral case is shown in (b), while (c) and (d) show the charged case without and with SCPC correction, respectively. Solid and dashed lines correspond to spin-up and spin-down states, respectively. Green lines denote defect-related states and the red line is the Fermi level.

As it is well known,[15] and as can be seen in Table 2 (or Fig.S8) too, the formation energy of a defect in a slab calculation never converges with the thickness of the vacuum region if no charge correction is applied. The reason for that is the fact that, due to the even distribution of the countercharge in the whole model and the localization of the actual charge in the solid part, the center of weight of the two charges will have an increasing distance, adding an ever increasing dipole interaction to the energy. This is corrected in the *a posteriori* methods, but the effect of the potential in the vacuum part still has its effect on the one-electron states. Fig.1 shows the changes in the band structure of a hydrogen-saturated (001) diamond slab upon incorporation of a neutral or negatively charged $N_CV_C$ center. As it is well known,[37] the neutral defect introduces three occupied and three unoccupied states into the gap. (At the chosen slab size, these levels split into defect bands, due to the interaction of the repeated defects.) When a negative charge is added, some of that is spilling into the vacuum (see Fig.S13.a), due to the potential of the countercharge there. The latter induces spurious image states, partially overlapping with the defect states, causing an erroneous occupation (Fig.1c). Using SCPC, this does not happen, because the potential is corrected self-consistently (Fig.S13.b). The band structure is reconstructed and the defect bands are clearly distinguishable (Fig.1d). The formation energy converges rapidly with vacuum thickness (Fig.S12), while *a posteriori* corrections do not yield reasonable results, since a substantial part of the charge is fully delocalized.

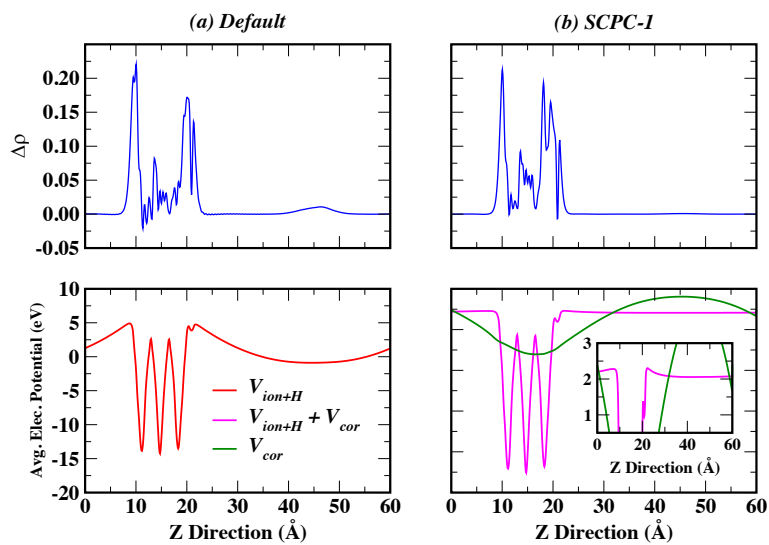

**Fig.2.** Planar average of the extra charge and of the electrostatic potential along the surface normal for an $O_2(-)$ molecule on the surface of an anatase-$TiO_2$ (001) slab, without (a) and with (b) the SCPC correction. In (b) the green curve shows the corrective potential and the inset in (b) shows a magnification of the Coulomb-tail due to the negative charge on $O_2$, which is not removed by the correction.

The hydrogenated diamond (001) surface has a negative electron affinity, so the charge spilling described above occurs for any vacuum thickness. However, this happens also for surfaces with positive electron affinity, as the vacuum thickness increases. We consider here an absorbed $O_2(-)$ molecule on the anatase-$TiO_2$ (101) surface. To study desorption of the molecule, a vacuum thickness of about 40 Å is needed (to have the minimally required 20 Å from the repeated surfaces[40] in case of the desorbed molecule in the vacuum). Fig.2 shows the distribution of the extra charge and of the electrostatic potential (Hartree + ionic contributions) without correction, and as obtained from SCPC. As can be seen, there is a spurious potential well in the vacuum, attracting charge spill-out. (For the given thickness of the solid part, charge spilling may occur even for a vacuum thickness below 30 Å.) Actually, this potential gives rise to "ghost states" (two-dimensional Rydberg states) in the middle of the vacuum (Fig.S14.c), and in a plane-wave calculation these get erroneously occupied (Fig.S15). Applying SCPC takes care of the problem (Fig.S14.d), while *a posteriori* energy correction methods cannot deal with such cases at all. We note that SCPC only corrects for the artifact in the potential but does not wash out the Coulomb-potential of the charge on the surface (see inset in Fig.2b).

In summary, we have devised a self-consistent potential correction (SCPC) method for electronic structure calculations in periodic charged systems, and implemented it into the VASP package (implementation into Quantum Espresso[41] will soon follow). The method works for bulk material as well as for slabs and wires, resulting not only in the correction of the total energy but also correcting the one-electron energies and wave functions self-consistently (and without post-processing). At the first step of the self-consistent cycle SCPC reproduces the *a posteriori* correction of the total energy as provided by the FN(V) and SLABCC methods. SCPC has a special significance for low dimensional systems, where it prevents the formation of artificial states in the vacuum, which arise due to the potential of the applied counter charge.

**Acknowledgement**. The help of G. Kresse, M. Schlipf and M. Marsman with the implementation of SCPC into the VASP code, as well as fruitful discussion with O. Andreussi and N. Marzari is greatly appreciated. The support of the DFG grant Nr. DE1158/8-1, the Research Training Group grant DFG-RTG2247, as well as the grant of the Supercomputer Center of Northern Germany (HLRN Grant No. hbc00027) is acknowledged. This work was also supported by the European Research Council (ERC-2015-AdG694097), the Cluster of Excellence 'Advanced Imaging of Matter' (AIM) and Grupos Consolidados (IT1249-19).

# References

[1] P. Deák, Th. Frauenheim, M. R. Pederson (eds.), *Computer Simulation of Materials at Atomic Level* [Wiley-VCH, Berlin 2000].

[2] A. Alkauskas, P. Deák, J. Neugebauer, A. Pasquarello, C. G. van De Walle (eds.), *Advanced Calculations for Defects in Materials* [Wiley-VCH, Berlin 2011].

[3] C. Freysoldt, B. Grabowski, T. Hickel, J. Neugebauer, G. Kresse, A. Janotti, and C.G. Van de Walle, Rev. Mod. Phys. **86**, 253 (2014).

[4] T.R. Durrant, S.T. Murphy, M.B. Watkins, and A.L. Shluger, J. Chem. Phys. **149**, (2018).

[5] T. Gake, Y. Kumagai, C. Freysoldt, and F. Oba, Phys. Rev. B **101**, 020102(R) (2020).

[6] C. A. Rozzi, D. Varsano, A. Marini, E. K. U. Gross and A. Rubio, Phys. Rev. B **73**, 205119 (2006).

[7] G. Makov and M.C. Payne, Phys. Rev. B **51**, 4014 (1995).

[8] S. Lany and A. Zunger, Phys. Rev. B **78**, 235104 (2008).

[9] C. Freysoldt, J. Neugebauer, and C.G. Van de Walle, Phys. Rev. Lett. **102**, 016402 (2009).

[10] H.-P. Komsa, T. T. Rantala, and A. Pasquarello, Phys. Rev B **86**, 045112 (2012).

[11] S.E. Taylor and F. Bruneval, Phys. Rev. B **84**, 075155 (2011).

[12] S.T. Murphy and N.D.M. Hine, Phys. Rev. B **87**, 094111 (2013).

[13] R. Rurali and X. Cartoixà, Nano Lett. **9**, 975 (2009).

[14] J.-Y. Noh, H. Kim, and Y.-S. Kim, Phys. Rev. B **89**, 205417 (2014).

[15] H.-P. Komsa and A. Pasquarello, Phys. Rev. Lett. **110**, 095505 (2013).

[16] H.-P. Komsa, N. Berseneva, A. V. Krasheninnikov, and R. M. Nieminen, Phys. Rev. X **4**, 031044 (2014); ibid. **8**, 039902(E) (2018).

[17] C. Freysoldt and J. Neugebauer, Phys. Rev. B **97**, 205425 (2018).

[18] M. Farzalipour Tabriz, B. Aradi, T. Frauenheim, and P. Deák, Comput. Phys. Commun. **240**, 101 (2019).

[19] R. Sundararaman and Y. Ping, J. Chem. Phys. **146**, 104109 (2017).

[20] S. Lany and A. Zunger, Phys. Rev. B **81** 113201 (2010).

[21] W. Chen and A. Pasquarello, Phys. Rev. B **88**, 115104 (2013).

[22] A.Y. Lozovoi and A. Alavi, Phys. Rev. B **68**, 245416 (2003).

[23] M. Otani and O. Sugino, Phys. Rev. B **73**, 115407 (2006).

[24] J. Xiao, K. Yang, D. Guo, T. Shen, H-X. Deng, S.-S. Li, J-W. Luo, and S. Wei, Phys. Rev. B **101**, 165306 (2020).

[25] N. Moll, Y. Xu, O.T. Hofmann, and P. Rinke, New J. Phys. **15**, 083009 (2013).

[26] N.A. Richter, S. Sicolo, S. V. Levchenko, J. Sauer, and M. Scheffler, Phys. Rev. Lett. **111**, 045502 (2013).

[27] M. Topsakal and S. Ciraci, Phys. Rev. B **85**, 045121 (2012).

[28] G. Kresse and J. Haffner, Phys. Rev. B **49**, 14251 (1994); G. Kresse and J. Furthmüller, *ibid*. **54**, 11169 (1996); G. Kresse and D. Joubert, Phys. Rev. B **59**, 1758 (1999).

[29] S. Lany and A. Zunger, Phys. Rev. B **78**, 235104 (2008).

[30] M.H. Naik and M. Jain, Comput. Phys. Commun. **226**, 114 (2018).

[31] C. Freysoldt, J. Neugebauer, and C.G. Van de Walle, Phys. Stat. Sol. (b) **248**, 1067 (2011).

[32] J.C. Womack, L. Anton, J. Dziedzic, P.J. Hasnip, M.I.J. Probert, and C.K. Skylaris, J. Chem. Theory Comput. **14**, 1412 (2018).

[33] L. Anton, J. Womack, and J. Dziedzic, (2013).

[34] G. Fisicaro, L. Genovese, O. Andreussi, N. Marzari, and S. Goedecker, J. Chem. Phys. **144**, (2016).

[35] R.D. Budiardja and C.Y. Cardall, Comput. Phys. Commun. **182**, 2265 (2011).

[36] I. Dabo, B. Kozinsky, E. Singh-Miller, and N. Marzari, Phys. Rev. B **77**, 115139 (2008); ibid. 84, 159910 (2011).

[37] M. Kaviani, P. Deák, B. Aradi, T. Frauenheim, J.-P. Chou, and A. Gali, Nanoletters **14**, 4772–4777 (2014).

[38] P.E. Blöchl, Phys. Rev. B **50**, 17953 (1994).

[39] J. P. Perdew, K. Burke, and M. Ernzerhof, Phys. Rev. Lett. **77**, 3865 (1996).

[40] P. Deák, J. Kullgren, B. Aradi, T. Frauenheim, and L. Kavan, Electrochim. Acta **199**, 27 (2016).

[41] P. Giannozzi, *et al.* J. Phys.: Condens. Matter. **29**, 465901 (2017).